\documentstyle[12pt,lathuile]{article}

\begin{document}

\title{ 
SEARCH FOR SUPERSYMMETRY, EXTRA DIMENSIONS \\
AND EXOTIC PHENOMENA AT LEP
}

\author{
Gabriella P\'asztor\footnotemark[1]
\\
{\em CERN, Gen\`eve 23, CH-1211, Switzerland} \\
}

\maketitle

\baselineskip=14.5pt

\begin{abstract}

The latest results on searches for supersymmetry, extra dimensions
and  exotic phenomena from the LEP collaborations are presented. 
No significant signal-like excess is observed in the data. The
results are interpreted in various models and robust constraints
are placed.

\end{abstract}

\renewcommand{\thefootnote}{\fnsymbol{footnote}}
\footnotetext[1]{Now at University of California, Riverside. 
On leave of absence from KFKI RMKI, Budapest.} 
\renewcommand{\thefootnote}{\arabic{footnote}}

\baselineskip=17pt

\newpage

\section{Introduction}

The Standard Model (SM) accurately describes the observed
phenomena  but leaves several fundamental questions unanswered.
Many extensions of the SM have been developed to solve these
puzzles. The LEP collider with its multi-purpose detectors ALEPH,
DELPHI, L3 and OPAL provided an excellent environment to explore 
possibilities beyond the SM both through direct searches for new
processes and through precise measurements of electroweak (EW)
parameters. Since 1996 when the centre-of-mass energy first reached
the WW production threshold, 700 pb$^{-1}$ of data per experiment
were collected at $\sqrt{s}=161-210$ GeV, of which 130 pb$^{-1}$
were recorded at $\sqrt{s}>206$ GeV.

In the following the latest LEP results are summarised on the various
flavours of supersymmetric models, theories with extra  dimensions and
a few selected topics from the rich field of exotic phenomena. 
For each model the phenomenological framework, the search
strategy and the achieved results (constraints at the 95\% confidence
level) are briefly discussed. 

In most cases the experiments provide limits on
the production cross-section of the studied processes with minimal model
assumptions, which are then interpreted within the framework of a given
model to derive constraints on the model parameters, particle masses.
Where available, the combined results of the four LEP experiments,
labelled by ADLO, are presented.

\section{\bf Supersymmetry}

Supersymmetry (SUSY) is a particularly promising extension of
the SM being theoretically well-motivated and also very successful from
the phenomenological point of view.

For each SM particle chirality state SUSY predicts a superpartner
differing in spin by half a unit. If SUSY were an exact symmetry the
particles and their superpartners would be degenerate in mass, thus SUSY
must be broken. Traditionally two theoretical scenarios are
examined\footnote{Recently a third scenario, anomaly-mediation, also
gains popularity.}: gravity-mediated and gauge-mediated SUSY breaking.
Both mechanisms assume that SUSY is broken in a \textit{hidden} sector
and SUSY breaking is transmitted from there to the \textit{visible}
sector where the SM and SUSY particles (sparticles) live. In models with
gravity mediated SUSY breaking (supergravity) the visible and hidden
sectors are coupled via gravitational interaction, while in models with
gauge mediated SUSY breaking (GMSB) the hidden sector couples to a
\textit{messenger} sector which in turn couples to the visible sector
via gauge interactions. 

SUSY fields can mix, thus the interaction eigenstates can differ 
from the mass eigenstates. The mixing of left- and right-handed 
scalar fermions (L-, R-sfermions, $\tilde{\mathrm f}_{\mathrm L},
\tilde{\mathrm f}_{\mathrm R}$)
%
%
is proportional to the corresponding fermion mass and is  negligible
for the first two generations. The  fermionic partners of the weakly
interacting gauge and Higgs\footnote{The Higgs sector of the SM needs
to be expanded to accommodate SUSY; in the Minimal Supersymmetric
extension of the SM (MSSM) two complex scalar Higgs doublets are
required leading to five Higgs bosons (h, H, A, H$^\pm$).} bosons form
six mass eigenstates: the charged higgsino and wino states give two
charginos ($\tilde\chi_i^\pm$), and the neutral bino, wino and
higgsino states give four neutralinos ($\tilde\chi^0_j$), where the
indices $i,j$ are ordered by increasing mass. The fermionic partners
of the strongly interacting gluons called gluinos ($\tilde g$) do not
mix with other states. 

\subsection{Supergravity}

In the most general case, MSSM has more than a hundred parameters in
addition to the SM ones. They include the couplings in the
superpotential and the masses and couplings in the soft SUSY breaking
terms.

In a constrained framework of MSSM (CMSSM), also called minimal
supergravity, the soft SUSY breaking parameters take a simple form at
the Planck scale: the scalar squared masses and the scalar couplings are
flavour diagonal and universal. Taking also the more general prediction
of the unification of gaugino masses, the number of parameters in the
soft SUSY breaking term can be reduced to four: the common scalar mass
($m_0$), scalar trilinear coupling ($A_0$), gaugino mass ($m_{1/2}$) and
the bilinear coupling of Higgs fields ($B_0$). $B_0$ can be exchanged to
the ratio of the vacuum expectation values (v.e.v.'s) of the Higgs
fields ($\tan\beta$) and $m_{1/2}$ to the SU(2) gaugino  mass parameter
at the EW scale ($M_2$).\footnote{The assumption of gaugino mass
unification at the GUT scale leads to $M_1 = 5/3 \tan^2\theta_W \, M_2$
for the U(1) gaugino mass parameter.} The mass parameters should not
exceed $\cal O$(TeV) so that SUSY remains a solution to the naturalness
problem.

In addition to the coupling $\mu$ of the Higgs fields, the
superpotential can also contain $R$-parity violating couplings
$\lambda_{ijk}, \lambda^\prime_{ijk}, \lambda^{\prime\prime}_{ijk}$
and $\mu_i^\prime$, where $i,j,k$ are generation indices. $R$-parity
is a multiplicative quantum number which takes the value of +1 for
SM particles and $-$1 for their superpartners. If $R$-parity is
conserved the constrained model can be described by only five extra
parameters. 

\subsubsection{\it $R$-parity conserving MSSM}
\label{sec:RPC}

The assumption of $R$-parity conservation has a crucial impact on
supersymmetric phenomenology. It implies that sparticles are always
produced in pairs and decay through cascade processes to SM particles
and to the lightest supersymmetric particle (LSP), which is stable. If
the LSP is neutral and weakly interacting, as favoured by cosmological
considerations, it escapes detection, resulting in sizable missing
energy.

All sparticles are expected to be pair-produced at LEP via $s$-channel
$\gamma$ or Z exchange. For third generation sfermions the production
cross-section depends on the mixing between the left- and right-handed
fields. $\tilde{\mathrm e}, \tilde\nu_{\mathrm e},
\tilde\chi_1^\pm$ and $\tilde\chi_i^0$ pair-production has $t$-channel
contribution, as well, and their cross-section strongly depends on the
model parameters.

SUSY phenomenology is largely determined by the nature of the LSP
and the next-to-LSP (NLSP). 
The LSP is usually considered to be
the lightest neutralino
(or the sneutrino). Accordingly, the following processes are searched
for:
\begin{list}{$\bullet$}{\itemsep=0pt \topsep=5pt \leftmargin=\parindent}
\item {\bf Chargino:}
$\tilde\chi^+_1 \tilde\chi^-_1 
\rightarrow 
(\tilde\chi^0_1 {\mathrm f \bar{f}'}) \,
(\tilde\chi^0_1 {\mathrm \bar{f} f'}) $
with $\tilde\chi^\pm_1$ decaying via $\tilde\chi^0_1$W$^\pm$ or
${\mathrm \tilde f \bar{f}^\prime}$;
\\
$\tilde\chi^+_1 \tilde\chi^-_1 
\rightarrow 
\tilde\nu \ell^+ \bar{\tilde\nu} \ell^-$;
\item {\bf Neutralino:}
$\tilde\chi^0_1 \tilde\chi^0_2 
\rightarrow 
\tilde\chi^0_1 \, (\tilde\chi^0_1$f\=f)
and
$\tilde\chi^0_2 \tilde\chi^0_2 
\rightarrow 
(\tilde\chi^0_1 {\mathrm f\bar{f}}) \,
(\tilde\chi^0_1 {\mathrm f\bar{f}})$
with $\tilde\chi^0_2$ decaying via $\tilde\chi^0_1$Z, 
$\tilde\chi^0_1$h/A or ${\mathrm \tilde f \bar f}$;
\item {\bf Sleptons:} 
$\tilde\ell^+ \tilde\ell^- 
\rightarrow (\tilde\chi^0_1 \ell^+) \, (\tilde\chi^0_1 \ell^-)$;
\item {\bf Light squarks and sbottom:} 
$\tilde{\mathrm q} \bar{\tilde{\mathrm q}} 
\rightarrow (\tilde\chi^0_1$q) $\, (\tilde\chi^0_1$\=q);
\item {\bf Stop:} 
$ {\mathrm \tilde t_1 \bar{\tilde t}_1 
\rightarrow (\tilde\chi^0_1 q) \, (\tilde\chi^0_1 \bar q)}$ 
via loop diagram with q=c,u;
\\
$ {\mathrm 
\tilde t_1 \bar{\tilde t}_1 
\rightarrow (\tilde\chi^+_1 b) \, (\tilde\chi^-_1 \bar b)
\rightarrow (\tilde\nu \ell^+ b) \, 
(\bar{\tilde\nu} \ell^- \bar b)}$. 
\end{list}

The chargino production cross-section is large except if
$\tilde{\nu}_{\mathrm e}$ is light and the destructive interference
between $s$- and $t$-channel processes becomes important. In this
case the search for neutralino production improves our sensitivity
for SUSY. If sfermions are heavy ($m_0>500$ GeV),
$\tilde\chi^\pm_1$ and $\tilde\chi^0_2$ decay dominantly via a W
and a Z boson, respectively.

The event properties depend significantly on the mass difference 
($\Delta M$) between the pair-produced sparticle and the LSP. In
the chargino search, for example, the case of 200 MeV $< \Delta M <
3$ GeV is treated separately using a dedicated analysis of events
with initial state radiation, and for $\Delta M <$ 200 MeV events
with tracks displaying kinks or impact parameter offsets and events
with heavy stable charged particles are studied. In CMSSM low 
$\Delta M$ is expected in the Higgsino region ($|\mu|<<M_2$); a
lower limit on the chargino mass of 92.4 GeV is obtained
independent of $\Delta M$ by the ADLO combination\cite{chargino,RPC}.

\begin{figure}[tp!]
\vspace{6.0cm}
\includegraphics{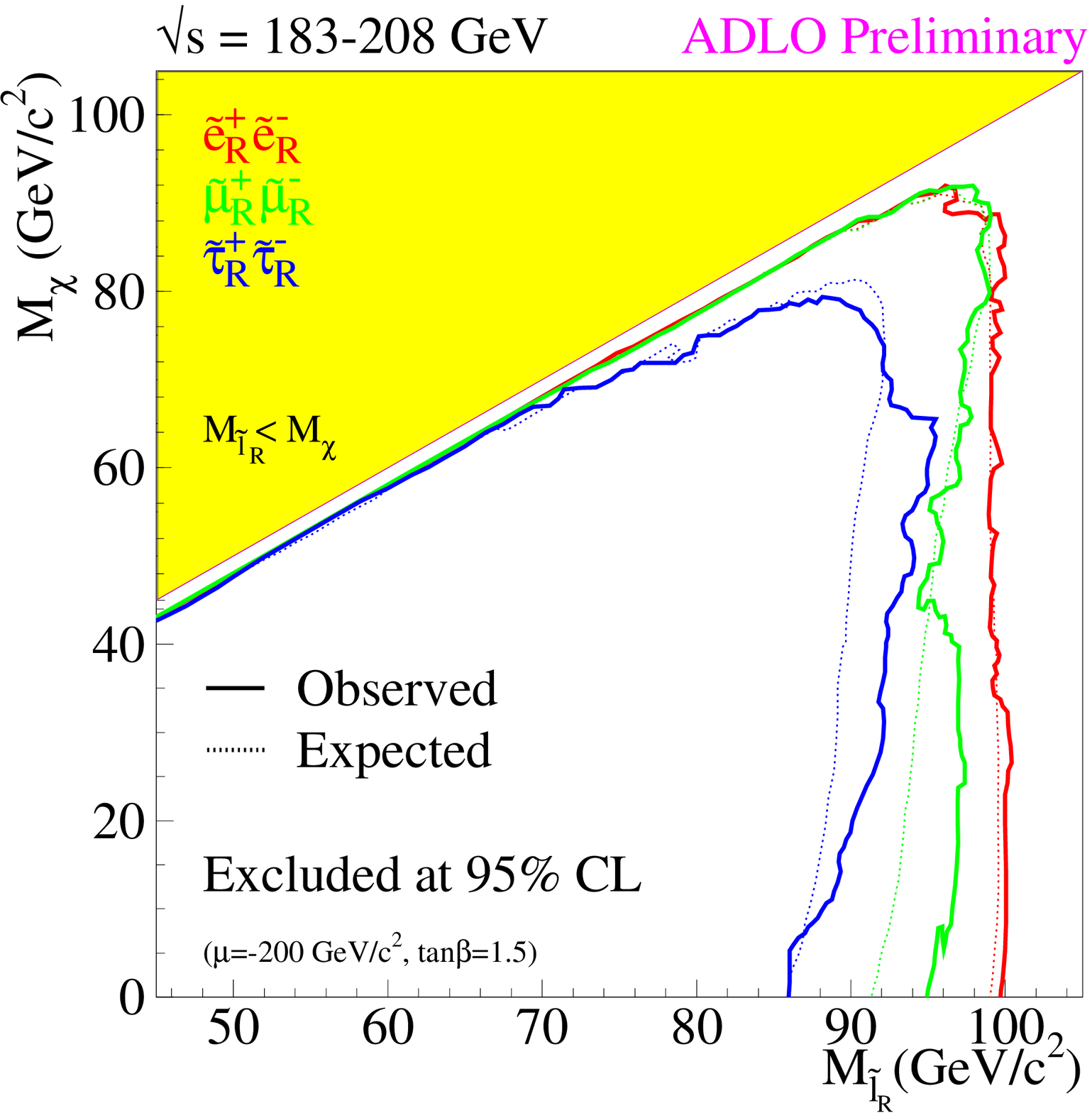}
\includegraphics{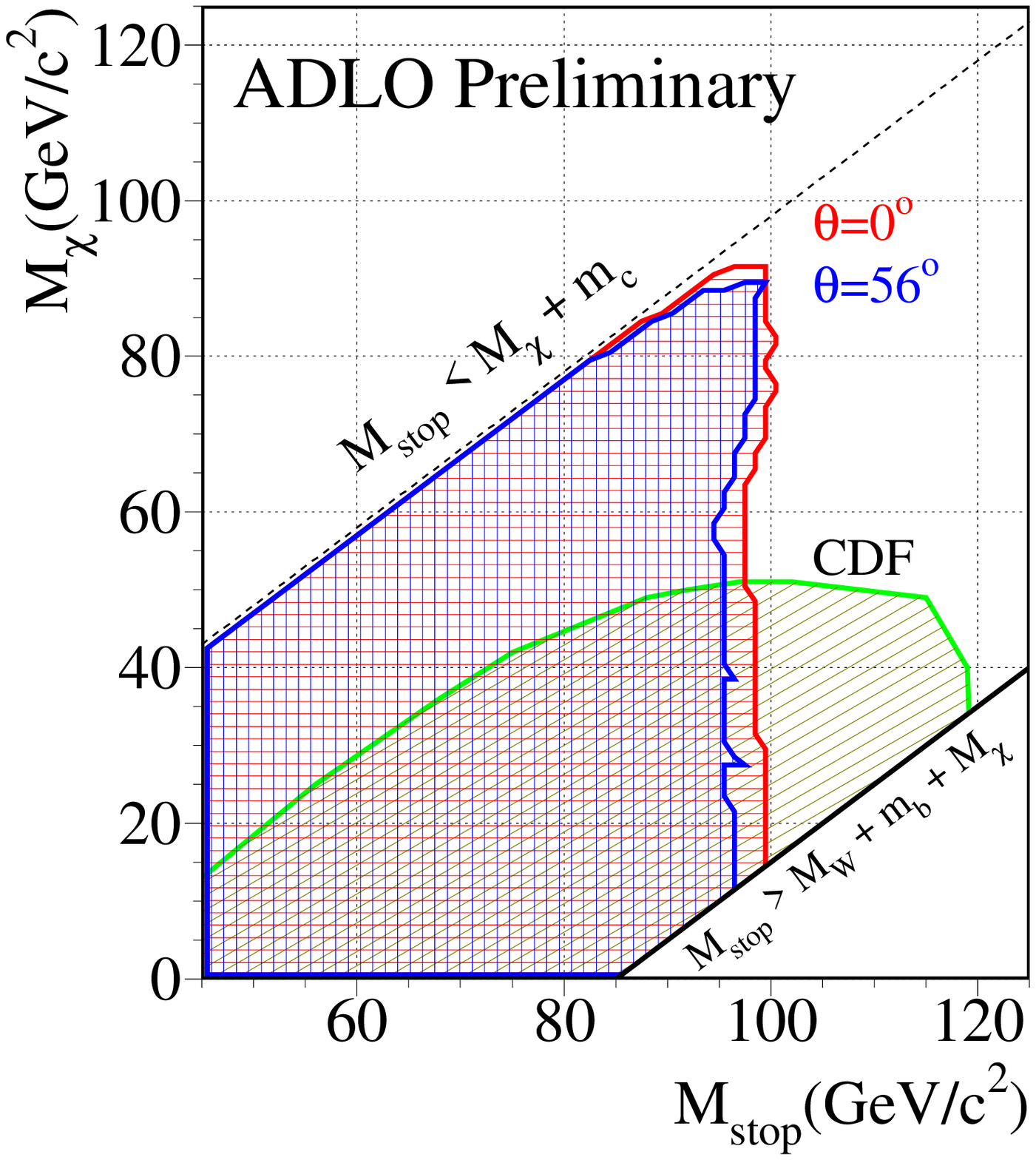}
\includegraphics{squark.ps}
\caption{\it
(left) Expected and observed exclusion domains in the 
$\tilde{\ell}_{\mathrm R} - \tilde\chi^0_1$ mass plane
taking into account the 
$\tilde\ell_{\mathrm R} \rightarrow \ell \tilde\chi^0_1$ 
branching ratio. 
(middle) Excluded domains in the 
$\tilde{\mathrm t}_1 - \tilde\chi^0_1$ mass plane
for no mixing and for the mixing angle giving the smallest
cross-section.
(right) Excluded domains in the $\tilde{\mathrm q} - 
\tilde{\mathrm g}$ mass plane assuming five degenerate 
$\tilde{\mathrm q}$ flavours.
\label{fig:sfermion} }
\end{figure}

The cross-section for $\tilde{\ell}_{\mathrm R}$ is smaller than for
$\tilde{\ell}_{\mathrm L}$, therefore $\tilde{\ell}_{\mathrm L}$ is
usually assumed to be out of the reach for the experiments and results
for $\tilde{\ell}_{\mathrm R}$ are given. For staus mixing may be 
sizable: the mass limits\cite{sleptons} shown on fig.\ref{fig:sfermion}
worsen by a few GeV when the Z boson is decoupled ($\theta_{\tilde\tau}
= 52^\circ$).

Using the standard searches for chargino and slepton production,
and developing dedicated analyses for $\tilde{\mathrm e}_{\mathrm
L}^+  \tilde{\mathrm e}_{\mathrm R}^-$ to cover the small $\Delta
M$ region and for $\tilde\chi^0_1 \tilde\chi^0_3$ in the so-called
\textit{corridor} where the chargino and the sneutrino are
degenerate in mass, the ALEPH collaboration set absolute lower
limits on the $\tilde{\mathrm e}_{\mathrm R}, \tilde{\mathrm
e}_{\mathrm L}$ and $\tilde\nu_{\mathrm e}$ masses\cite{Asleptons}
of 73, 107 and 84 GeV, respectively, assuming sfermion and gaugino
mass unification and no sfermion mixing. Including the constraints from
neutral Higgs boson searches, the results on the selectron masses
can be improved to 77 and 115 GeV for a top mass of 175 GeV. Within
CMSSM for $A_0 = 0$ the obtained bounds are 95, 152 and 130 GeV,
respectively. 

If kinematically allowed, the $\tilde {\mathrm t}_1 \rightarrow
\tilde\nu \ell^+$b decay mode is dominant over $\tilde\chi^0_1$c. The
combined LEP results\cite{squarks} exclude stop and sbottom masses
up to 94$-$100 GeV for $\Delta M > 10$ GeV depending on the
search channel and the mixing angle. If charginos and sleptons are
light, four-body stop decays $ {\mathrm \tilde t_1 \bar{\tilde t}_1 
\rightarrow (\tilde\chi^+_1 b) \, (\tilde\chi^-_1 \bar b)}$ $ {\mathrm
\rightarrow (\tilde\chi^0_1  f \, \bar{f}' \, b)  \, (\tilde\chi^0_1 
\bar{f} \, f' \, \bar b)}$ can be enhanced, leading to less stringent
limits on the stop mass\cite{Asquarks} than those on
fig.\ref{fig:sfermion}.

ALEPH performed a search for stop production with small $\Delta M$
looking for charged particle tracks with significant lifetime. The
obtained absolute mass limit is 63 GeV independent of the values of
$\Delta M, \mu$ and $\tan\beta$ explored in the scan\cite{Asquarks}. 

The searches for acoplanar jets can be translated into constraints
shown in fig.\ref{fig:sfermion} on mass degenerate squarks (left-
and right-handed ${\mathrm \tilde u, \tilde d, \tilde s, \tilde c,
\tilde b}$) within the MSSM with lowest order GUT relations between
the soft SUSY-breaking gaugino mass terms\cite{squarks}.

\begin{figure}[tp!]
\vspace{7.2cm}
\includegraphics{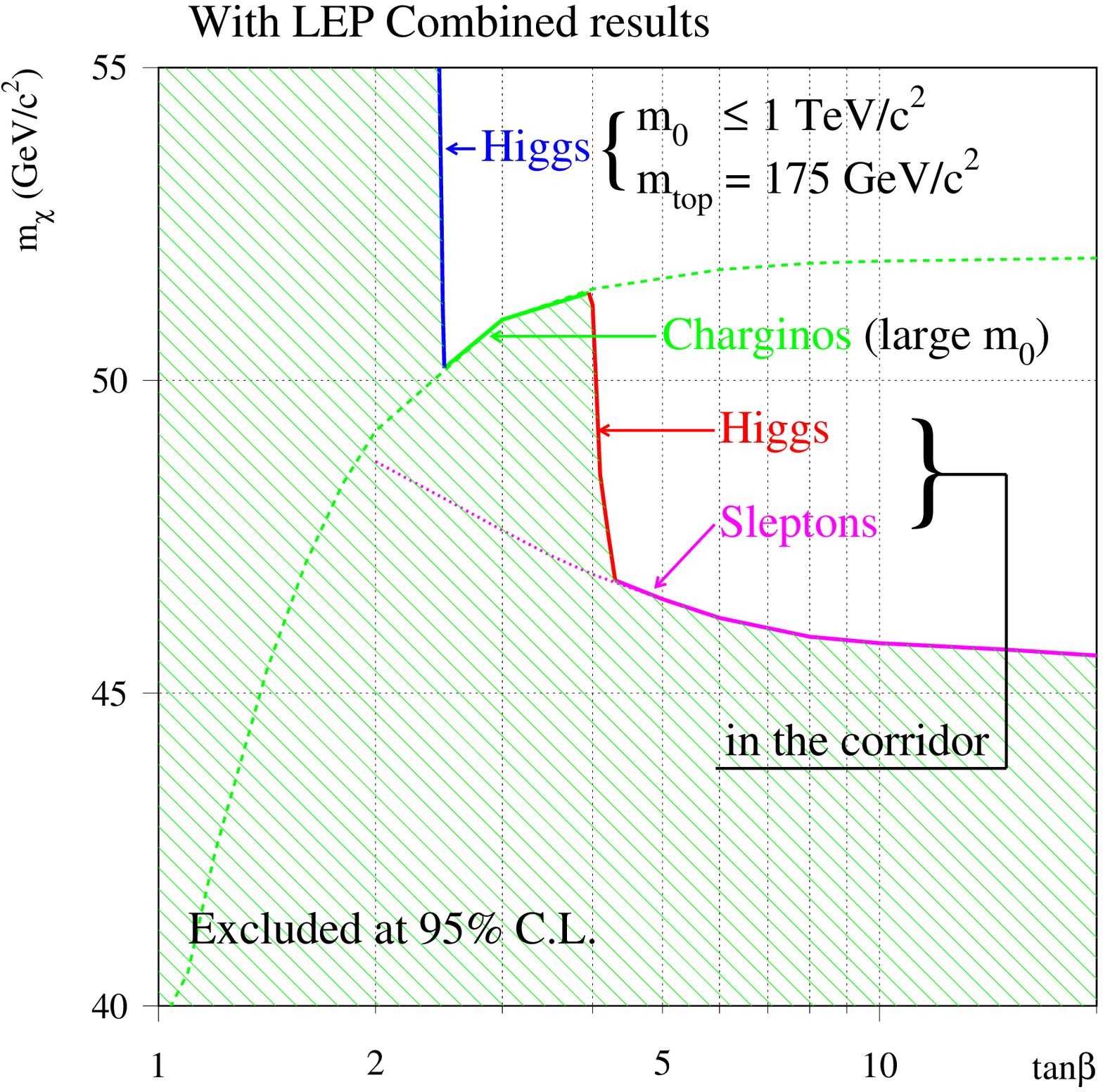}
\includegraphics{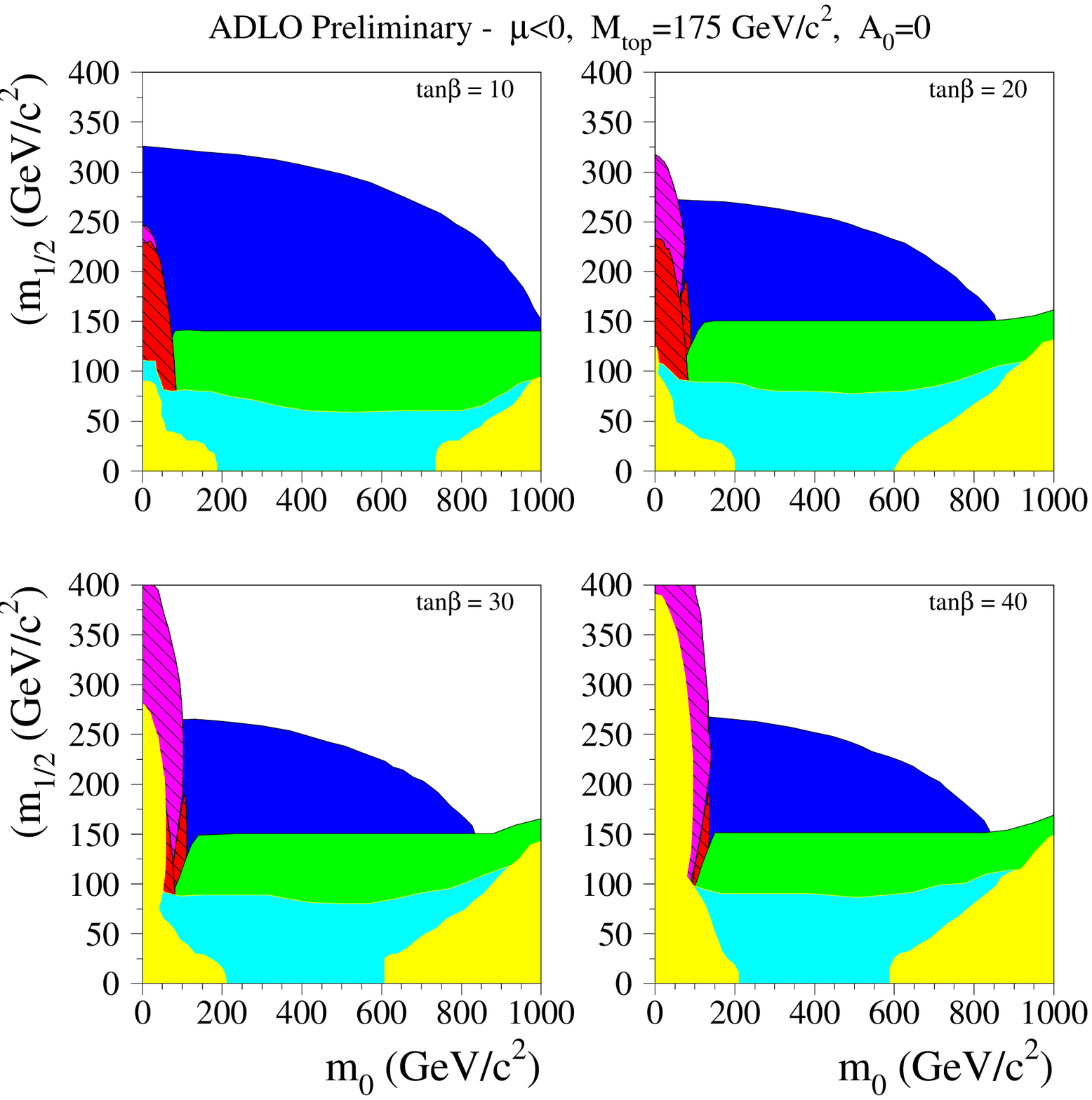}
\caption{\it
(left) Lower limit on the $\tilde\chi^0_1$ LSP
mass. The results at low $\tan\beta$ depend strongly on the top
mass via the Higgs constraint. 
The theoretical uncertainty is $\cal O$(1 GeV) due to the use
of tree level gaugino masses and lowest order relations for gaugino
unification. 
(right) Excluded regions in CMSSM in the $m_0 - m_{1/2}$ plane by
no CMSSM solution (yellow), constraints from LEP1 (light blue),
chargino (green), standard slepton (red), standard
hZ (dark blue), neutralino stau cascade (brown),
heavy stable stau (magenta) searches.
\label{fig:neutralino}
}
\end{figure}

When combining the negative results of chargino, neutralino, slepton and
Higgs boson searches, limits on the $\tilde\chi^0_1$ LSP mass can be
obtained as shown in fig.\ref{fig:neutralino} assuming gaugino and
sfermion mass unification at the GUT scale and negligible mixing in the
stau sector\cite{neutralino}. Stau mixing may lead to scenarios with
mass degenerate stau and LSP and weaken the derived limits at large
$\tan\beta$. Dedicated searches for $\tilde\chi^\pm_1 \rightarrow
\tilde\tau \nu$ and $\tilde\chi^0_2 \rightarrow \tilde\tau \tau$ cover
this region. 

Within CMSSM, using also the constraints from the measurement of the Z
width and the searches for heavy stable stau production, the obtained
bounds on the parameters, see fig.\ref{fig:neutralino}, can be
translated into 52.0$-$59.0 GeV lower limits on the LSP mass\cite{CMSSM}
depending on sign\,$\mu$ and the top mass for $A_0=0$ and $m_0 < 1$ TeV.

\subsubsection{\it $R$-parity violating MSSM}

There are no theoretical or experimental arguments excluding
$R$-parity violation (RPV), moreover, the branching ratios of 
$R$-parity violating decay modes of sparticles can be comparable or
even larger than the $R$-parity conserving ones. If $R$-parity is
violated, sparticles can be singly produced and can decay directly
to SM  particles. Therefore, the predicted signatures differ from
the characteristic missing energy signature of $R$-parity
conserving processes. 

With the MSSM particle content, $R$-parity violating interactions
are described with a gauge-invariant superpotential that includes
the  following Yukawa terms\footnote{ $R$-parity could be
spontaneously broken through a $\tilde\nu$ acquiring a
non-zero v.e.v. This can be described by a bilinear term $\mu_i^\prime
L_i H_2$.}: 
$$
{W}_{RPV}  = 
    \lambda_{ijk}      L_i L_j {\overline E}_k
 +  \lambda^{'}_{ijk}  L_i Q_j {\overline D}_k
 +  \lambda^{''}_{ijk} {\overline U}_i {\overline D}_j 
    {\overline D}_k, 
\label{lagrangian}
$$
where $L$ and $Q$ are lepton and quark left-handed
doublet superfields, $\overline E$, $\overline D$ and
$\overline U$ are right-handed  singlet charge-conjugate
superfields for the charged  leptons, down- and up-type quarks,
respectively, and 
$i,j,k$ are the generation indices of the superfields.
$\lambda_{ijk}$ is non-vanishing only if $i < j$, 
and $\lambda^{''}_{ijk}$ is non-vanishing only for $j < k$, 
therefore there are a total of 45 $R$-parity violating Yukawa
couplings. 

It is usually assumed that the sparticles are pair-produced via
$R$-parity conserving processes described in sec.\ref{sec:RPC}.
Two different scenarios are then probed. In the first scenario,
called indirect decays, the decays of sfermions via the lightest
neutralino, $\tilde\chi^0_1$, are considered,  where
$\tilde\chi^0_1$ is treated as the LSP and assumed to decay via an
$R$-parity violating Yukawa coupling. In the second scenario, direct
decays of sparticles to SM particles are investigated.  In this
case, the sparticle is assumed to be the LSP, such that $R$-parity
conserving decay modes do not contribute. In both scenarios, it is
assumed that only one of the 45 Yukawa couplings is non-zero at a
time, motivated by constraints from low energy experiments.  It is
also assumed that the LSP decays promptly, implying  a very short
lifetime, and therefore a mass larger than 10 GeV for the lightest
neutralino.

The topologies resulting from RPV decays of pair-produced
sparticles are numerous and extremely varied: direct decays of
sfermions lead to 4-fermion, direct decays of charginos and
neutralinos to 6-fermion, indirect decays of sfermions to
8-fermion and finally indirect decays of charginos to 10-fermion
final states with almost any combination of species and flavours
of final state particles.

\begin{figure}[tp!]
\vspace{5.0cm}
\includegraphics{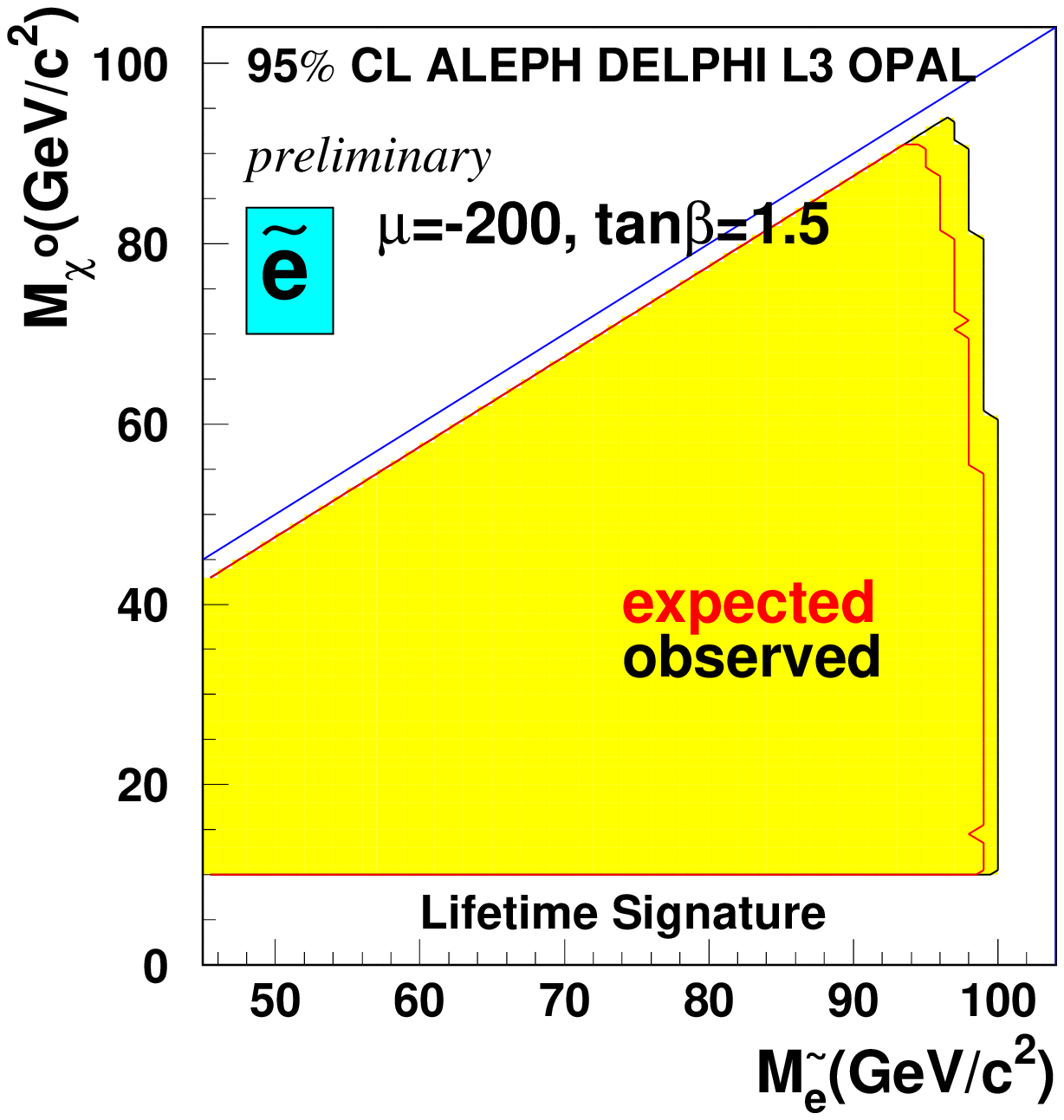}
\includegraphics{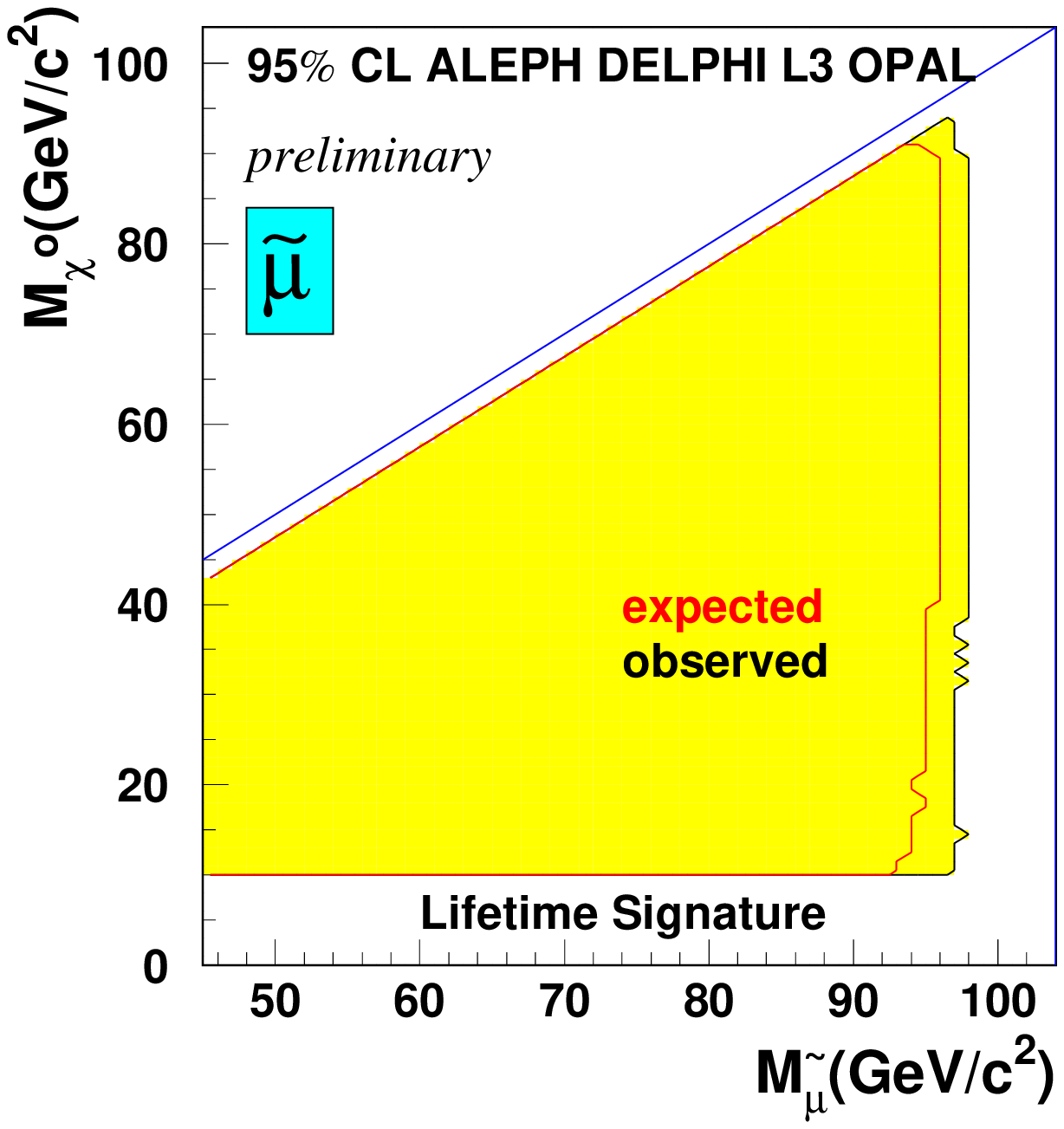}
\includegraphics{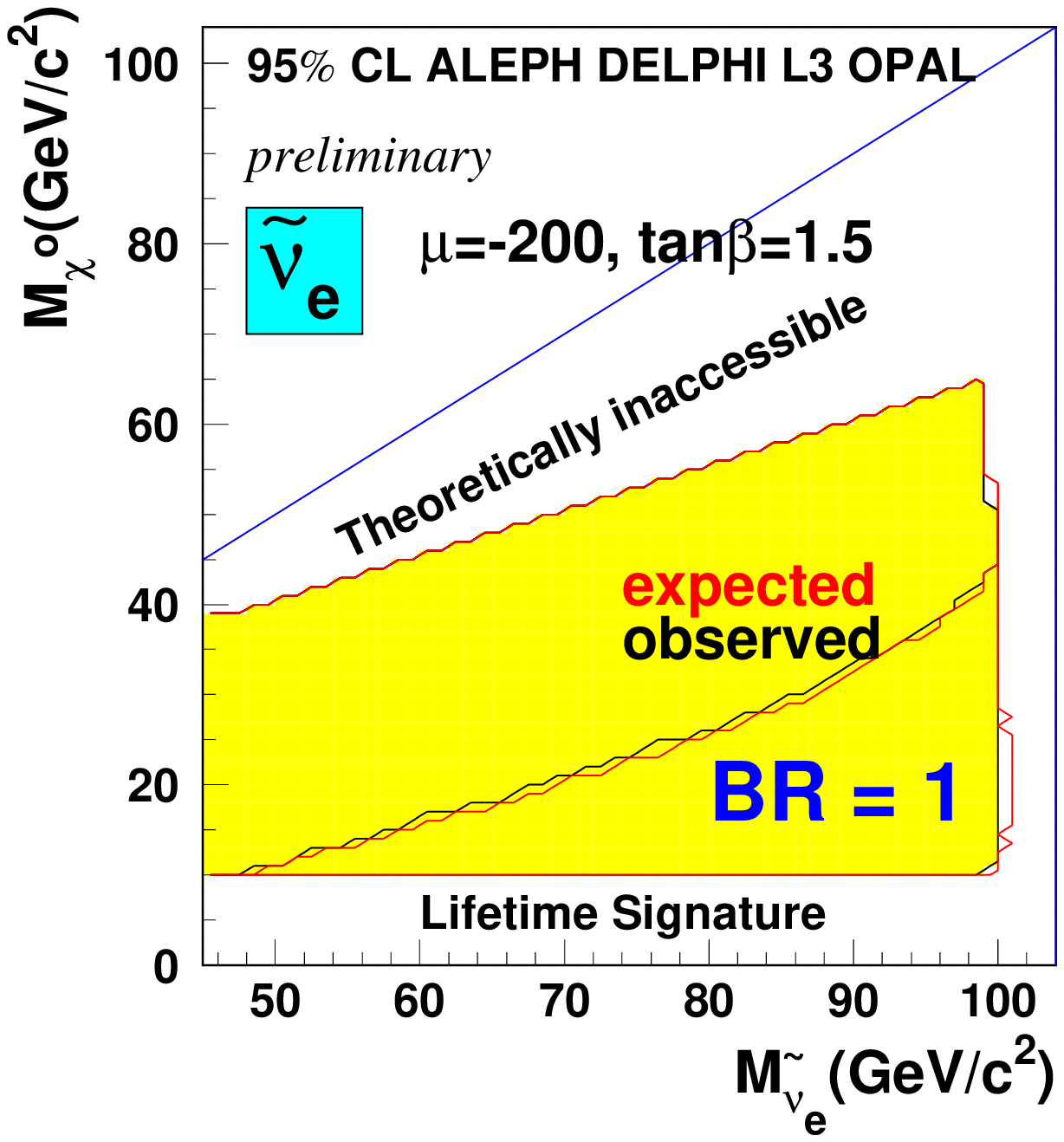}
\caption{\it Excluded regions in the $\tilde\ell (\tilde\nu) -
\tilde\chi_1^0$ mass plane from searches for indirect slepton 
decays via a $\lambda$ coupling in RPV CMSSM.
\label{fig:LLE}
}
\end{figure}

The first LEP combined results\cite{RPVsleptons} are shown in
fig.\ref{fig:LLE} for indirect decays of sleptons via $\lambda$
couplings.  In general the limits for $\lambda^\prime$ and
$\lambda^{\prime\prime}$ couplings\cite{RPVresults} are less stringent
due to the presence of more final state quark jets.

\begin{figure}[tp!]
\vspace{5.3cm}
\includegraphics{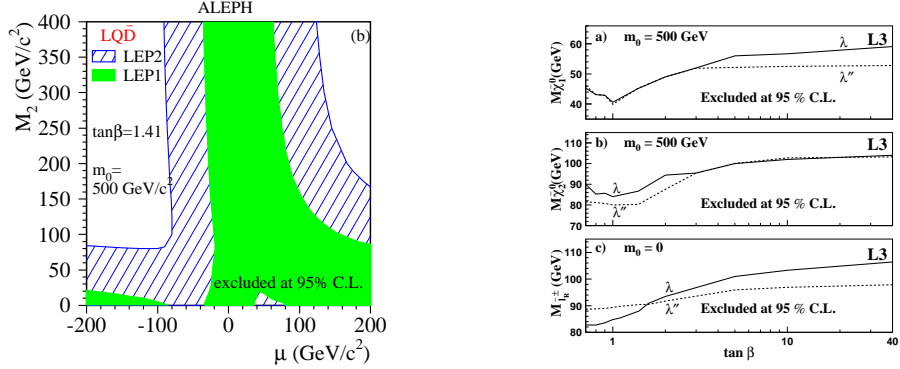}
\includegraphics{fig5.epsi}
\caption{\it (left) Excluded regions in the $\mu - M_2$ plane 
from searches for gaugino decays via $\lambda'$ coupling in
RPV CMSSM. (right) Absolute lower limits on the sparticles masses 
for $0 \leq M_2 \leq 1$ TeV and -0.5 TeV
$\leq \mu \leq$ 0.5 TeV. The indicated values of $m_0$ correspond
to the global minima on the mass limit.
\label{fig:mu-M2}
}
\end{figure}

The results of the chargino and neutralino searches are used to
constrain the $\mu - M_2$ plane for a given $m_0$ and $\tan\beta$ as
shown in fig.\ref{fig:mu-M2}. When combining all RPV searches absolute
lower limits are also derived on the sparticle masses from a scan of
the CMSSM parameter space.

By studying single $\tilde\nu$ production in the e$\gamma
\rightarrow \tilde\nu \ell$ process the ALEPH collaboration set
upper limits on $\lambda_{1jk}$ couplings assuming that the
sneutrinos are degenerate in mass. These results improve existing
limits from charged current universality for masses $M_{\tilde\nu}
< 190$ GeV\cite{RPVsnu1}. From the search for resonant $\tilde\nu$
production\cite{RPVsnu2} the DELPHI collaboration derived limits
on $\lambda_{1j1}$ couplings most stringent (few times
10$^{-3}$) for $\tilde\nu$ masses close to the LEP centre-of-mass
energies. 

\subsection{GMSB}

In the minimal version of GMSB, six new parameters are introduced in
addition to the SM parameters: the SUSY breaking scale ($\sqrt{F}$),
the messenger scale ($M$), the messenger index giving the number of
messenger particle sets ($N$), the mass scale which determines the
SUSY particle masses at the messenger scale ($\Lambda$), the ratio of
the v.e.v.'s of the two Higgs doublets ($\tan\beta$) and the sign of
the Higgs sector mixing parameter (sign\,$\mu$).

In GMSB, the SUSY partner of the graviton, the gravitino
($\tilde{\mathrm G}$), is expected to be the LSP with a mass,
typically less than 1 GeV, determined by $\sqrt{F}$. The NLSP is
either the lightest neutralino or a slepton. In the latter case two
possibilities are considered: a stau NLSP or slepton co-NLSPs when
all sleptons are light and  degenerate in mass. The experimental
signatures crucially depend on the NLSP decay length which can take
basically any value.

\begin{figure}[tp!]
\vspace{6.0cm}
\includegraphics{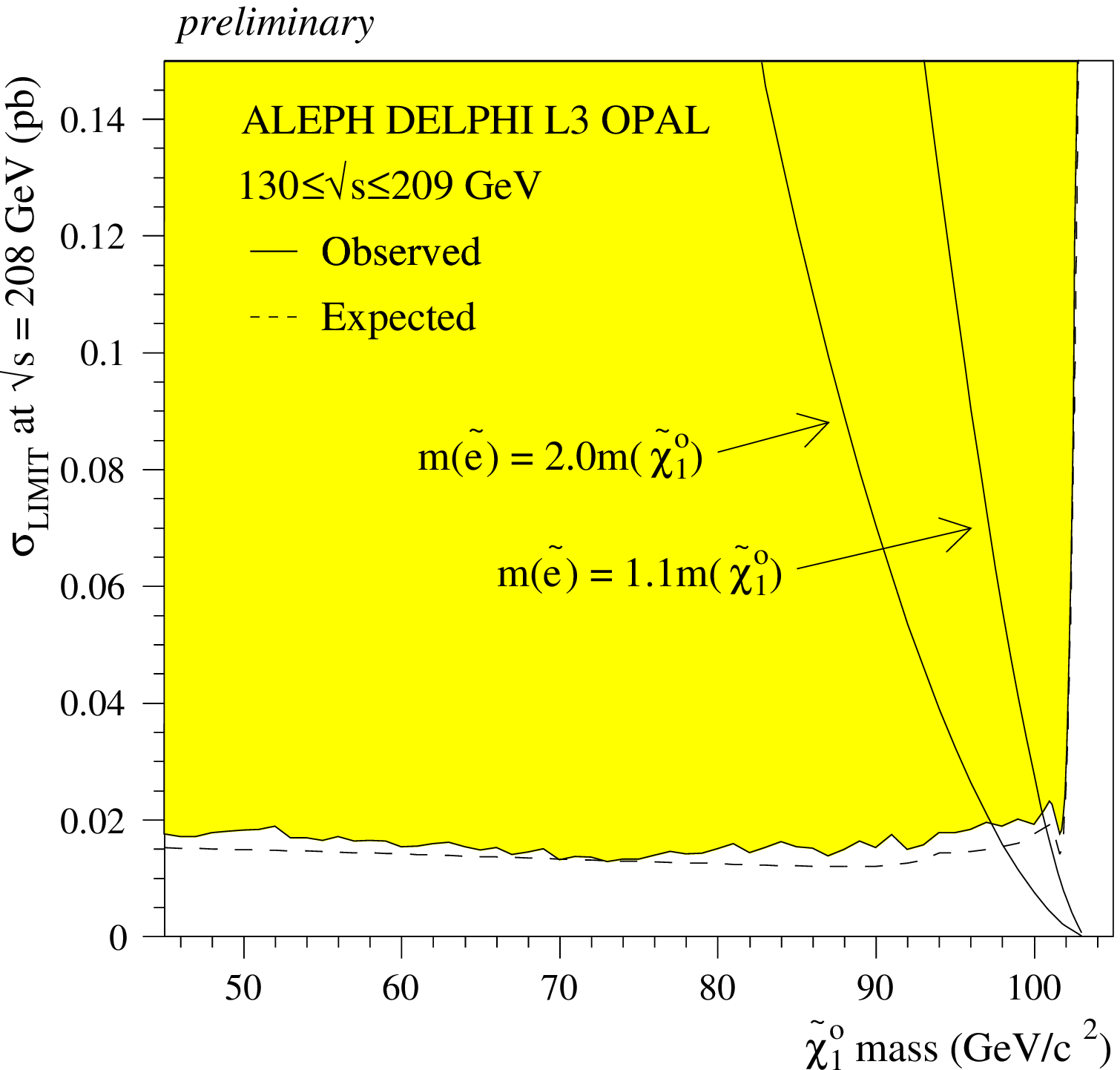}
\includegraphics{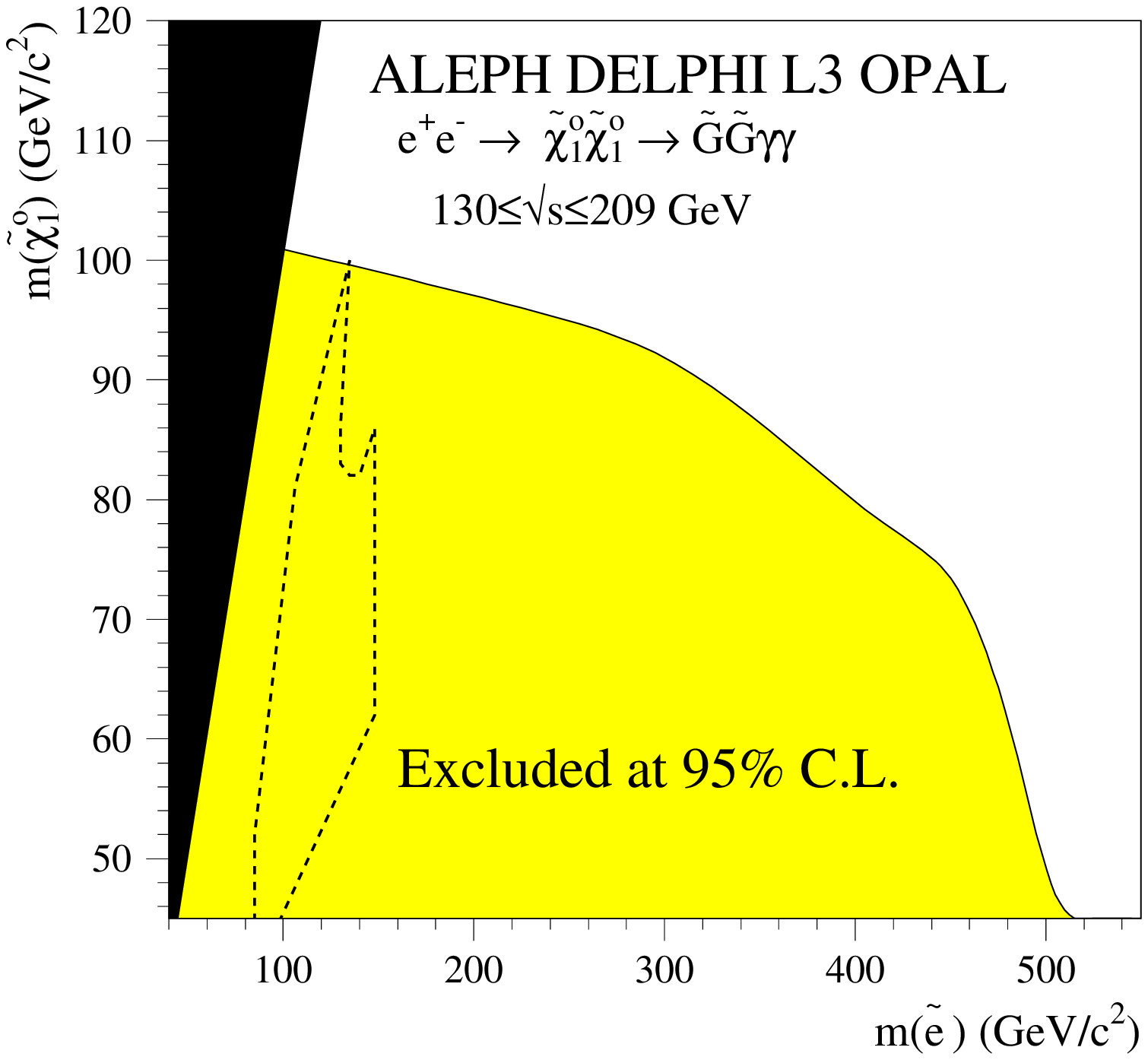}
\includegraphics{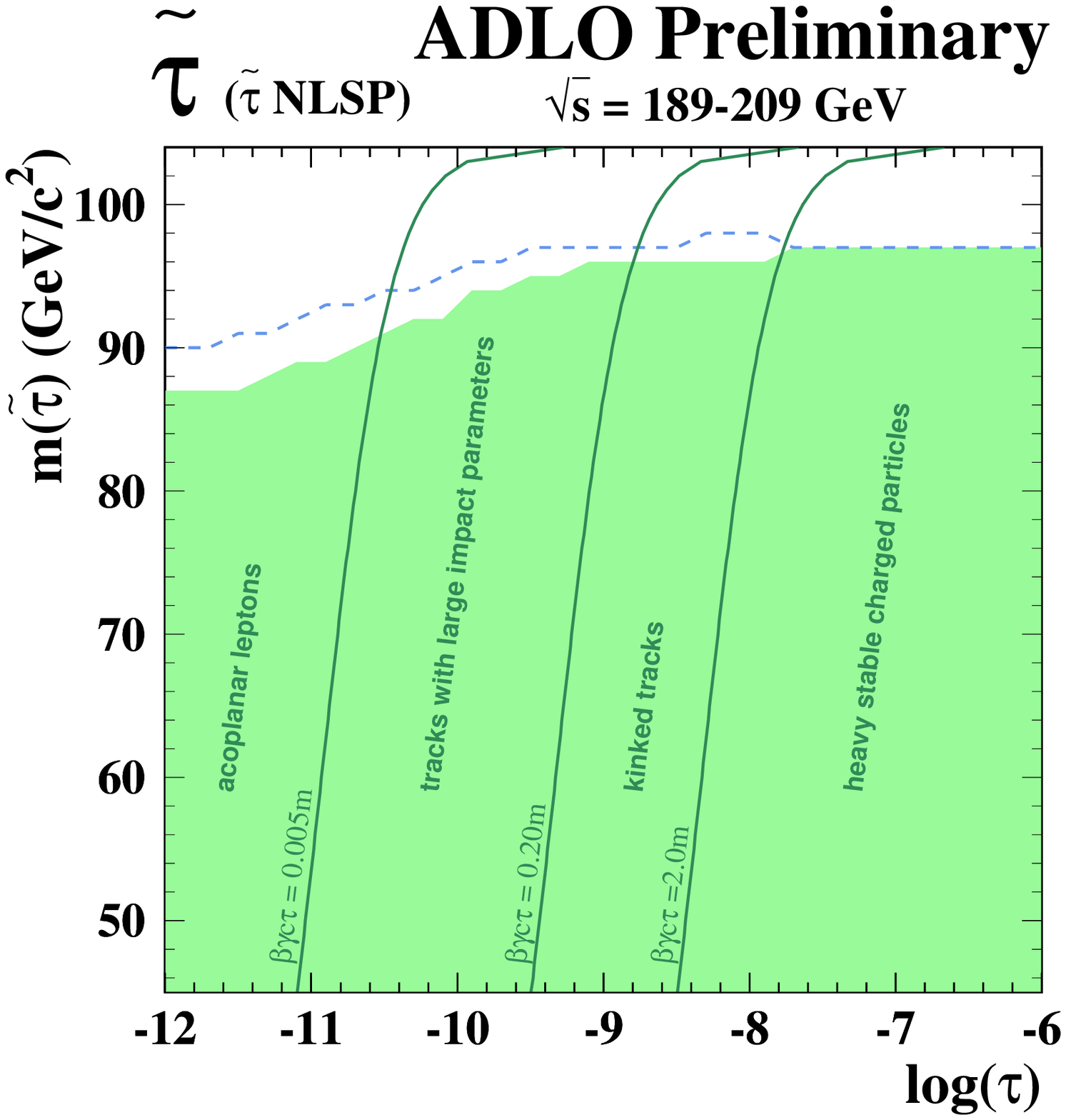}
\caption{\it
(left) Observed and expected cross-section limits for $\tilde\chi^0$ 
pair-production with theoretical cross-section curves for different
$\tilde{\mathrm e}$ masses in GMSB. 
(middle) Excluded region overlayed on the 
area consistent with the CDF ee$\gamma\gamma$ + missing $E_{\mathrm T}$
event. $\tilde{\mathrm e}_{\mathrm L}$ and 
$\tilde{\mathrm e}_{\mathrm R}$ are assumed to be degenerate in mass 
and the $\tilde\chi^0_1$ is assumed to be pure bino. 
(right) Observed and expected exclusion regions on the $\tilde\tau$
mass - lifetime plane, indicating the corresponding search 
topologies.
\label{fig:GMSB-xsec}}
\end{figure}

In the neutralino NLSP scenario neutralinos are either produced in pairs
directly or indirectly via slepton  ($\tilde\ell^+ \tilde\ell^-
\rightarrow$ $\tilde\chi^0_1 \ell^+$ $\tilde\chi^0_1 \ell^-$) and
chargino ($\tilde\chi^+_1 \tilde\chi^-_1 \rightarrow$  $\tilde\chi^0_1
{\mathrm W}^{+*}$ $\tilde\chi^0_1 {\mathrm W}^{-*}$) pair-production or
neutralino cascade decay  ($\tilde\chi^0_1 \tilde\chi^0_2 \rightarrow$ 
$ \tilde\chi^0_1$ $\tilde\chi^0_1 Z^{*}$), if the corresponding
sparticles are light. The lightest neutralino will decay into a
gravitino and a photon, giving topologies with photon pairs and missing
energy. Depending on its lifetime the photons are either originating
from the interaction point or have a large impact parameter. Indirect
production of neutralinos plays an important role if the neutralino
lifetime is long, and therefore the direct pair-production is
invisible.  In fig.\ref{fig:GMSB-xsec} the upper limit on the
cross-section of neutralino production is shown for prompt decays,
together with the excluded regions on the $\tilde\chi^0_1 -
\tilde{\mathrm e}$  mass plane\cite{GMSBphotons,photons}. Combining
searches for all lifetimes the ALEPH collaboration reports\cite{AGMSB} a
neutralino LSP mass limit of 54 GeV.

In the case of slepton NLSP, which is expected to decay into a
lepton and a gravitino, the events are characterised by leptons and
missing energy. In the slepton co-NLSP case the sleptons are either
pair-produced directly,  or through chargino  ($\tilde\chi^+_1
\tilde\chi^-_1  \rightarrow $ $\tilde\ell^+ \nu_\ell$ $\tilde\ell^-
\bar\nu_\ell$) and neutralino production ($\tilde\chi^0_1
\tilde\chi^0_1  \rightarrow$  $\tilde\ell^\pm \ell^\mp$ 
$\tilde\ell^\pm \ell^\mp $). The picture is slightly different in
the stau-NLSP scenario, where stau production can also happen
through $\tilde{\mathrm e}$ or $\tilde\mu$ pair-production
($\tilde\ell^+ \tilde\ell^-   \rightarrow $ $\tilde\chi^0_1 \ell^+$
$\tilde\chi^0_1 \ell^- \rightarrow $ $\tilde\tau^\pm \ell^\mp
\ell^+$ $\tilde\tau^\pm \ell^\mp \ell^-$). On fig.\ref{fig:GMSB-xsec}
the excluded regions are shown in the $\tilde\tau$ mass -- lifetime
plane from the searches for slepton pair-production with different
lifetimes\cite{GMSBleptons}. 


 
\begin{figure}[tp!]
\vspace{6.0cm}
\includegraphics{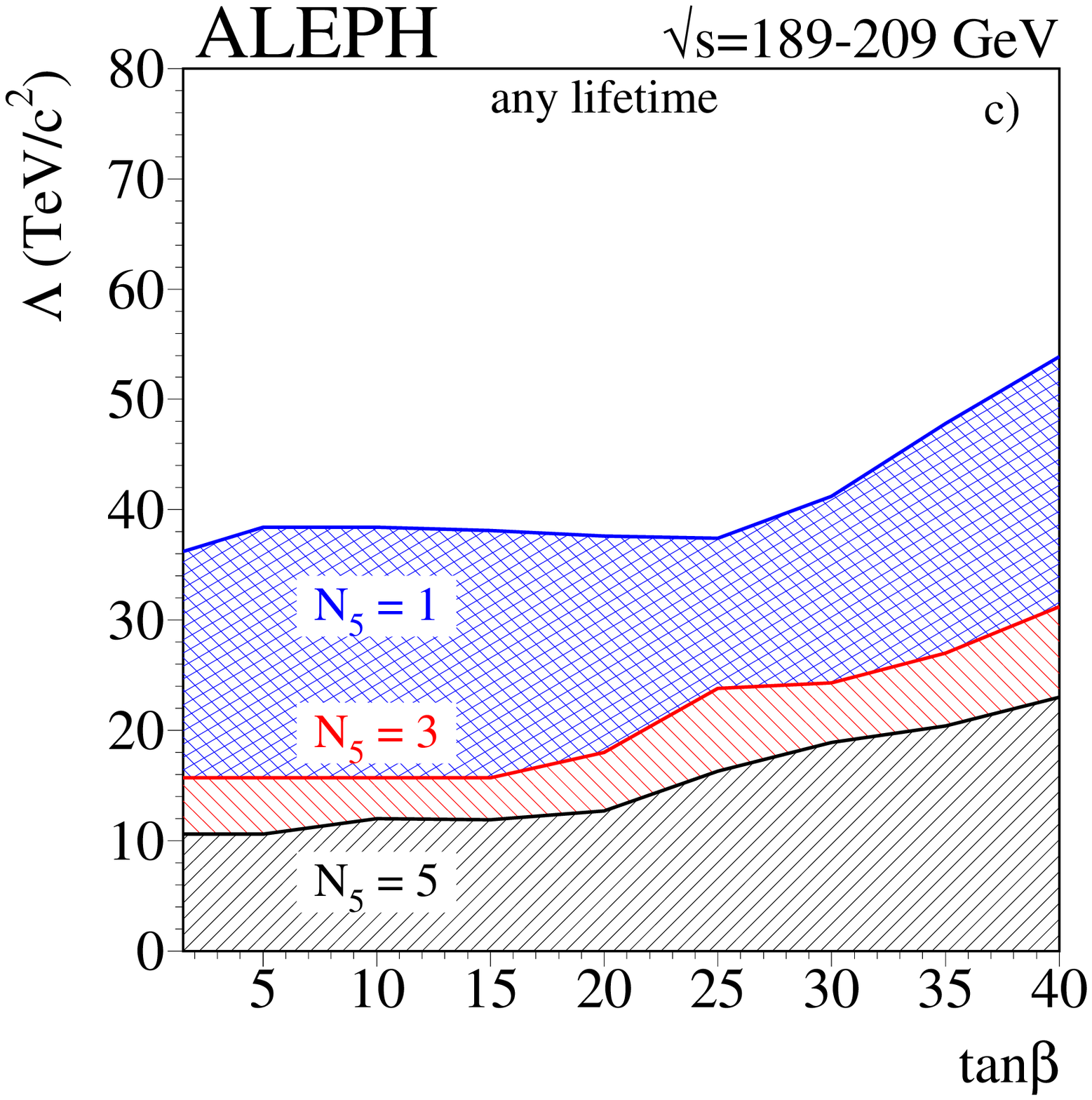}
\includegraphics{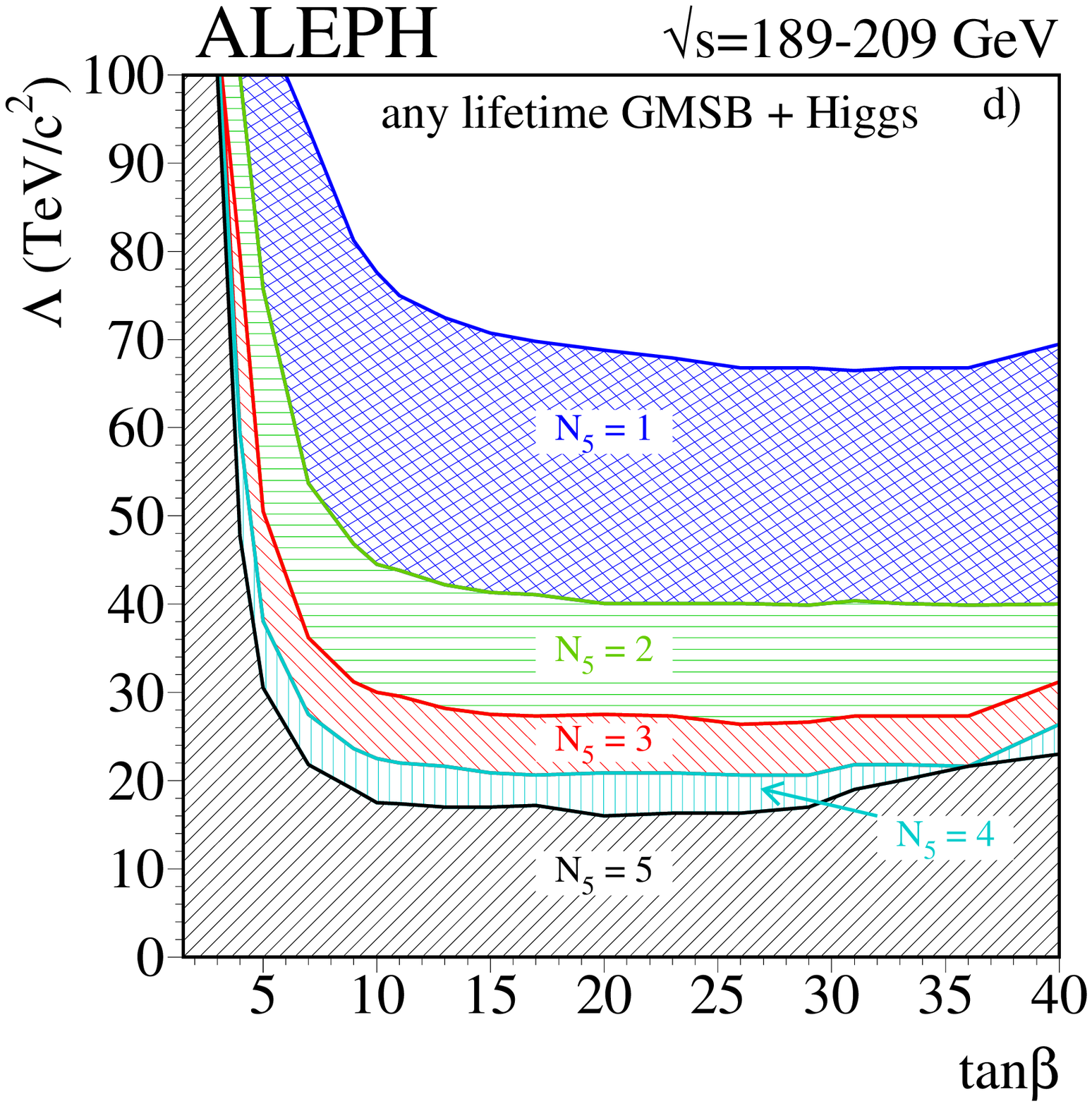}
\includegraphics{pn504_23_col.epsi}
\caption{\it
Excluded regions on the $\Lambda - \tan\beta$ plane (left) from  
GMSB searches and (middle) in combination with
Higgs searches. (right) Limits on sparticle masses from 
GMSB parameter scan.
\label{fig:GMSBscan}
}
\end{figure}

By scanning the GMSB parameter space constraints on the parameters and 
absolute limits on the sparticle masses can be derived. The excluded
regions on the $\Lambda - \tan\beta$ plane\cite{AGMSB} are shown in
fig.\ref{fig:GMSBscan} together with the obtained mass
limits\cite{OGMSB}.

\section{\bf Extra dimensions}

Models with extra dimensions have been introduced to solve the
hierarchy problem of the SM through geometrical considerations. The
original model of Arkani-Hamed--Dimopoulos--Dvali (ADD) of large
extra dimensions appeared in 1998 and triggered the
development of a vast number of new models.  

Most LEP results are derived in the ADD framework, which assumes
$n$ compact extra dimensions of size $R$, with the Planck scale,
$M_D$, in $D=4+n$ dimensions set close to the EW scale. SM
particles propagate in the usual four, while gravity in $D$
dimensions. The 4-dimensional Planck scale, $M_{\mathrm  Planck}$,
satisfies $M_{\mathrm  Planck}^2 \sim R^n M_D^{n+2}$.

The Klauza-Klein (KK) excitations of the graviton
($\mathrm{G_{KK}}$) couple to the momentum tensor and contribute to
most SM processes. The fermion- and boson-pair cross-sections are
modified   $$\sigma = \sigma_{\mathrm SM} + \alpha_{\mathrm G}
\sigma_{\mathrm int} + \alpha_{\mathrm G}^2 \sigma_{\mathrm
grav}$$  with  $\alpha_{\mathrm G} = \frac{2\lambda}{\pi}
M_{\mathrm S}^{-4}$.
$\lambda$ depends on the details of the model and it is usually set
either to $+1$ or $-1$ to allow for both positive and negative
interference. $M_{\mathrm S}$ is the ultraviolet cut-off scale close to
$M_{\mathrm D}$. The most stringent constraint,  
$M_{\mathrm S} > 1.18 / 1.17$ TeV for
$\lambda=+1 / -1$, comes from Bhabha
scattering\cite{ADDee}.  The combination of the results of the LEP
experiments is expected to only slightly improve the limits.  The
combined result from photon pair-production\cite{ADDgg-estar}
gives  $M_{\mathrm S} > 0.93 / 1.01$ TeV for $\lambda=+1 / -1$,
whereas the individual experiments placed lower bounds between 0.80
and 0.96 TeV.

\begin{figure}[tp!]
\vspace{5.5cm}
\includegraphics{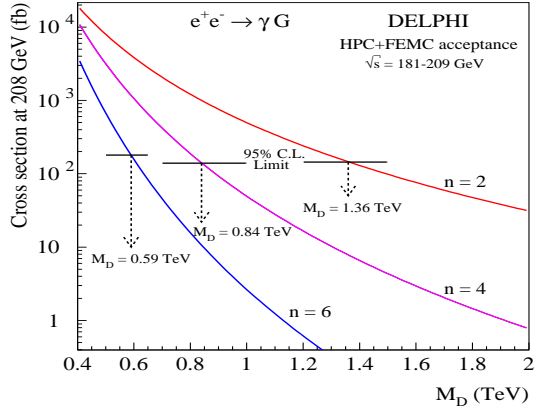}
\caption{\it Search for direct graviton production in ADD. Limits on
$M_{\mathrm D}$ can be turned into constraints on the size of the
extra dimensions giving $R < $0.25 mm, 13 pm, 54 fm for $n$=2,4,6,
respectively.
\label{fig:ADDdir}}
\end{figure}

The search for direct graviton production in the process
e$^+$e$^-$ $\rightarrow$ $\gamma G_{\mathrm KK}$ or  Z$G_{\mathrm
KK}$ is sensitive to the $D$-dimensional  Planck scale
itself\cite{photons}. The results for
different numbers of extra dimensions are shown in 
fig.\ref{fig:ADDdir}. 

\section{\bf  Exotic phenomena}

\subsection{Single top production}

At LEP single top production can be searched for
in several theoretical frameworks, such as flavour changing
neutral currents (FCNC), 4-fermion contact interactions or
R-parity violating SUSY\cite{singletop}. 

In the SM FCNC is forbidden at tree level in good agreement with
the observed low rates of such processes, thus all extensions of
the SM must face the challenge to sufficiently suppress FCNC. On
the other hand FCNC processes are ideal to look for new physics
due to the small SM background.

\begin{figure}[tp!]
\vspace{6.3cm}
\includegraphics{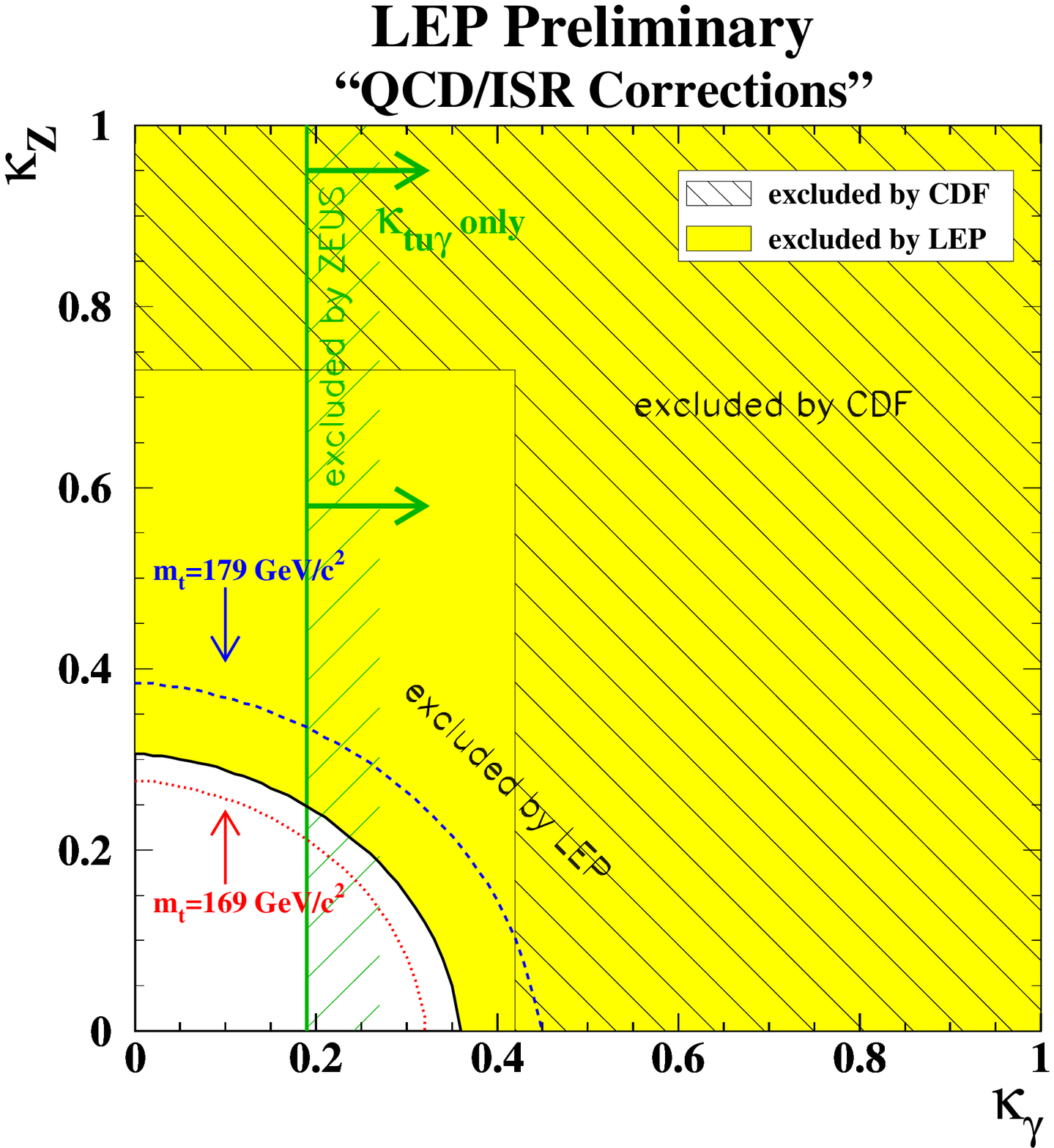}
\includegraphics{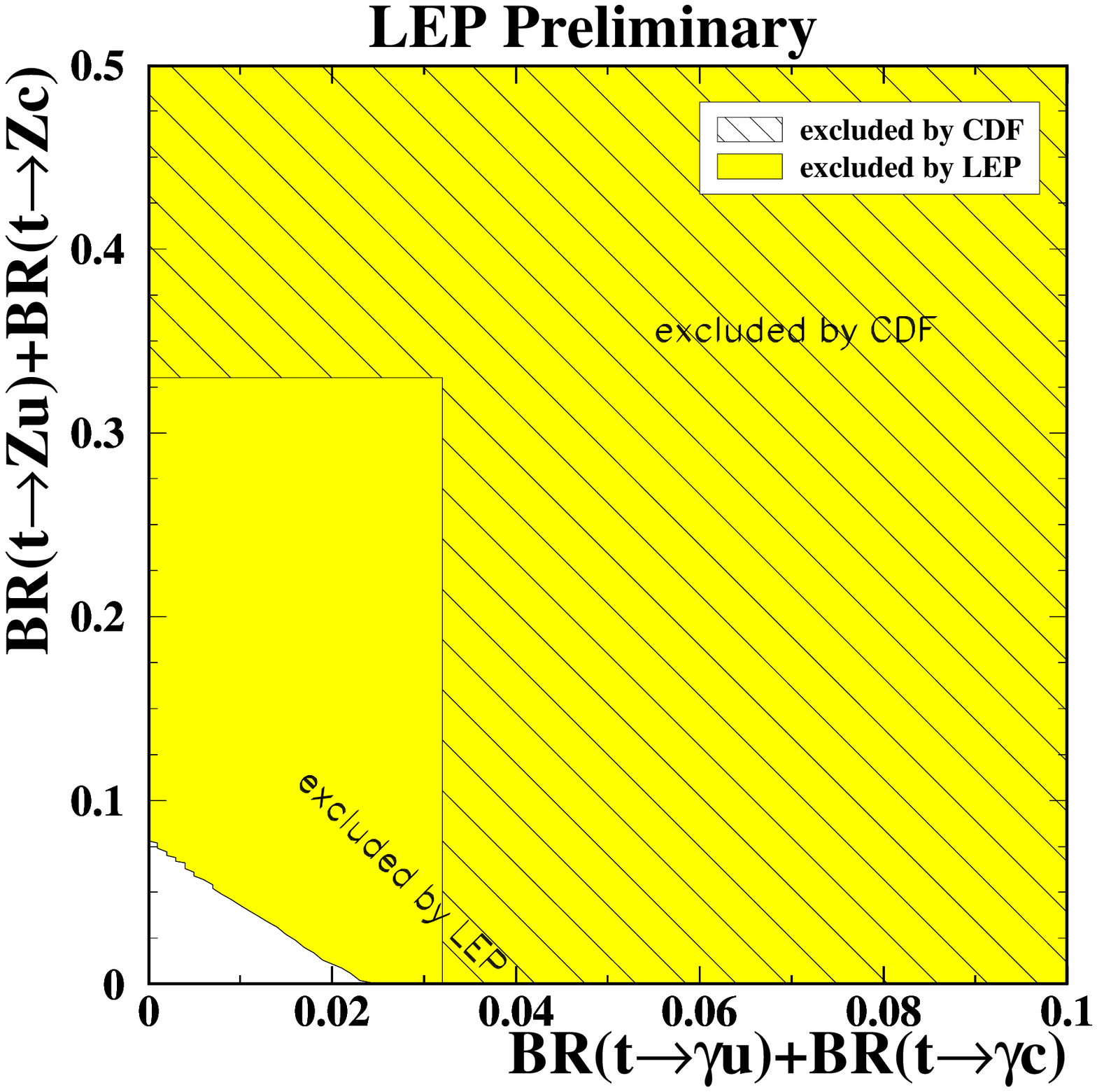}
\caption{\it Single top production via FCNC. Excluded regions 
in the (left) $\kappa_\gamma - \kappa_{\mathrm Z}$ and (right)
$Br({\mathrm t \rightarrow \gamma q}) - 
Br({\mathrm t \rightarrow Z q})$ planes 
\label{fig:singletop}}
\end{figure}

The amplitude of e$^+$e$^-$ $\rightarrow$ \=tc(u) via FCNC is
parametrised in terms of anomalous vertices with strength
$\kappa_{\mathrm Z}$ and $\kappa_\gamma$. The top quark
decay \=t $\rightarrow$ \=bW$^-$ would then lead to 4-fermion final
states of \=b$\ell^-\bar\nu_\ell$c(u) and \=bq\=q$^\prime$c(u). The
combined LEP results\cite{LEPsingletop} set a strong bound on
$\kappa_{\mathrm Z}$ and $\kappa_\gamma$ which can also be expressed 
as branching ratio limits as shown in fig.\ref{fig:singletop}.

\subsection{Exited leptons}

Models in which fermions have substructure at a scale $\Lambda$
attempt to explain the pattern of fermion generations. The
existence of exited states of the SM fermions would be natural in
such models.

Exited leptons can be produced in pairs or in association with a
SM lepton, both proceeding through $s$-channel $\gamma$ or Z
exchange. There is also a $t$-channel contribution for the first
generation. Exited leptons are expected to decay via the
emission of an EW gauge boson ($\gamma$, Z or W).

Searches for the pair-production process yield mass limits very
close to the kinematic limit. Single production provides a tool to
extend  the mass reach. The search channels include
\begin{list}{$\bullet$}{\itemsep=0pt \topsep=5pt \leftmargin=\parindent}
\item
$\nu^* \nu \rightarrow \gamma\nu\nu, 
{\mathrm W}\ell\nu, 
{\mathrm Z}\nu\nu$
\item
$\ell^* \ell \rightarrow \gamma\ell\ell, 
{\mathrm W}\nu\ell, 
{\mathrm Z}\ell\ell$ 
\end{list}
leading to widely different event topologies.

In the phenomenological models used at LEP the
couplings  V$\ell^*\ell$ associated to the two gauge groups SU(2)
$\times$ U(1) are proportional to the factors $f/\Lambda$ and
$f^{'}/\Lambda$, respectively. It is usual to set $|f|=|f^\prime|$
when deriving limits.

\begin{figure}[tp!]
\vspace{5.5cm}
\includegraphics{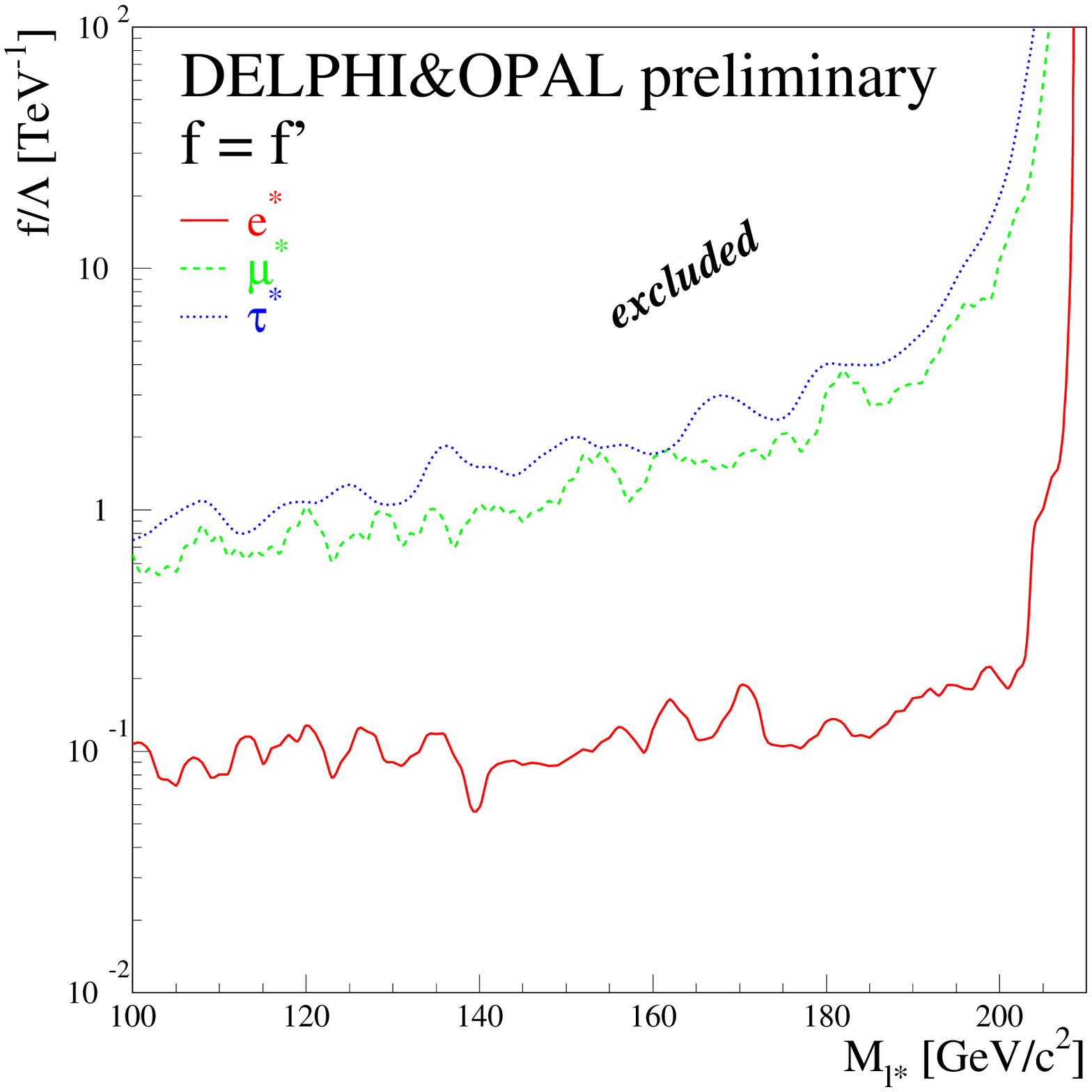}
\includegraphics{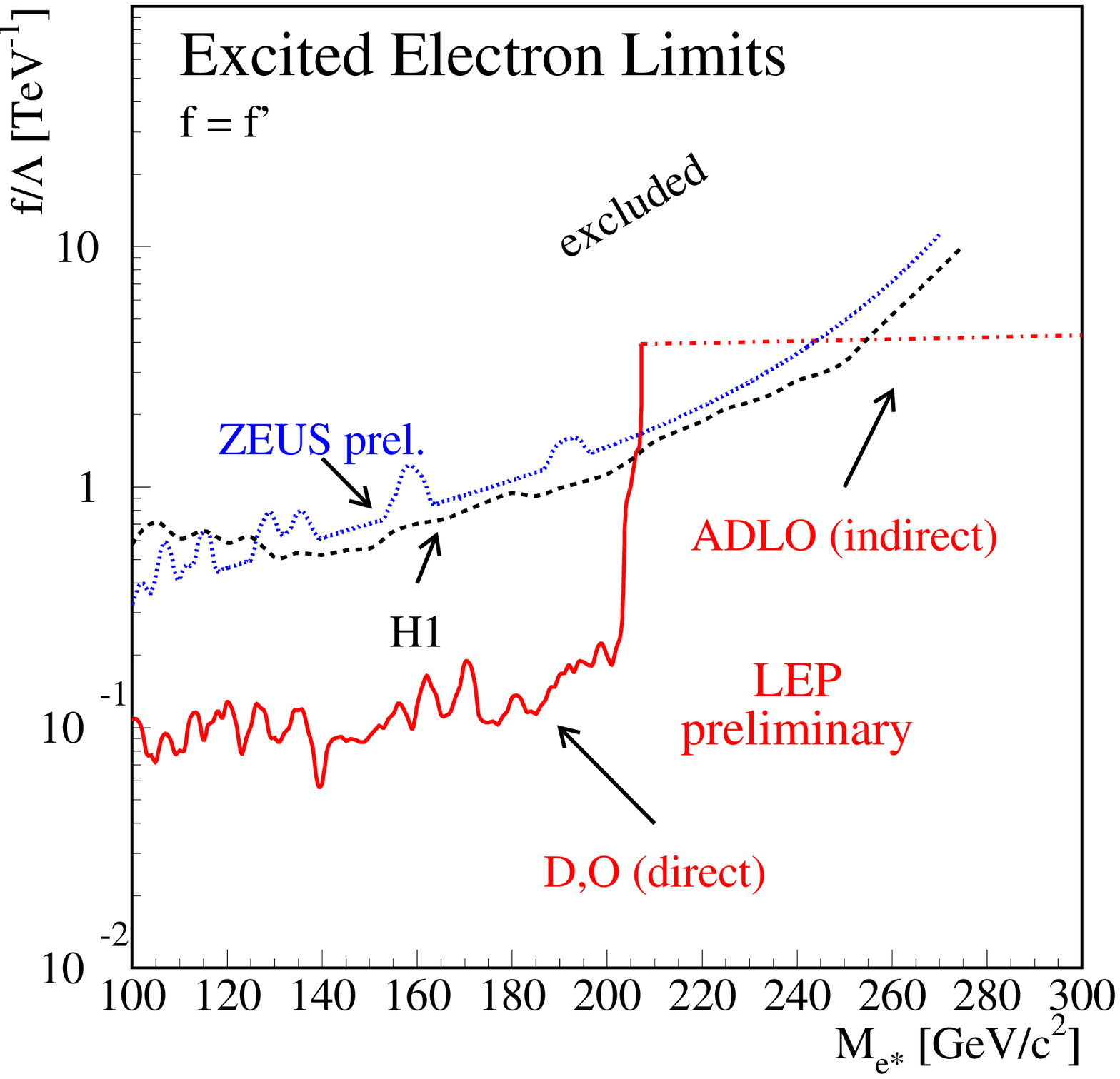}
\caption{\it Exited lepton production. lower bounds on the parameter
f/$\Lambda$ (left) for the three exited lepton flavours from direct
and (right) for exited electrons from  direct and indirect
searches.
\label{fig:lstar}}
\end{figure}

For exited electrons, if $f \neq -f^\prime$, the experimental reach
can be further increased by measuring the process e$^+$e$^-$ 
$\rightarrow \gamma\gamma(\gamma)$ which can have a contribution from
$t$-channel e$^*$ production. The latest results of the LEP
collaborations\cite{lstar} are shown in fig.\ref{fig:lstar}.

\subsection{Technicolor}

Technicolor (TC) solves the naturalness and hierarchy problems of
the SM by predicting new strong interactions which break
dynamically the EW symmetry without the presence of an elementary
Higgs scalar. Simple versions of TC disagree with the observations,
thus more and more refined proposals were born. The LEP searches
are guided by the Walking Extended TC (Straw Man 
Model). It predicts TC scalar and vector mesons,
$\pi_{\mathrm T}$ and  $\rho_{\mathrm T}$, which can be light
enough to be observed at LEP. 

OPAL looked for the process $\rho^0_{\mathrm T} \rightarrow
\pi^+_{\mathrm T} \pi^-_{\mathrm T}, \pi^0_{\mathrm T} \gamma$,
while
DELPHI also considered $\rho^0_{\mathrm T} \rightarrow
\pi^+_{\mathrm T} {\mathrm W}^-_{\mathrm L},
{\mathrm W}^+_{\mathrm L} {\mathrm W}^-_{\mathrm L}$ and 
$\rho^0_{\mathrm T} \rightarrow$ q\=q. Lower limits on the
techni-rho mass above 200 GeV have been set\cite{TC}.


\section{\bf Conclusion}

The LEP experiments explored all main areas and many corners for new
physics during the last years, but no significant deviation from the SM
has been found. In particular, the LEP constraints on SUSY are rather
robust for variations of the model. We should therefore continue to look
for the signs of a more fundamental theory at the TeV scales beyond LEP
reach at the next generation of colliders.

\section{\bf Acknowledgements}

I would like to thank the four LEP collaborations for providing their
latest results. I am grateful to \'A. Csilling for carefully reading
the manuscript.

\end{document}